# Relationship between the boson peak and first sharp diffraction peak in glasses

**Running title: Boson peak dynamics in glass**

Dan Kyotani[1,†], Soo Han Oh[1,†], Suguru Kitani[2], Yasuhiro Fujii[3,4], Hiroyuki Hijiya[5], Hideyuki Mizuno[6], Shinji Kohara[7], Akitoshi Koreeda[8], Atsunobu Masuno[9], Hitoshi Kawaji[2], Seiji Kojima[1], Yohei Yamamoto[1], Tatsuya Mori[1*]

[1]Department of Materials Science, University of Tsukuba, 1-1-1 Tennodai, Tsukuba, Ibaraki 305-8573, Japan

[2]Materials and Structures Laboratory, Institute of Integrated Research, Institute of Science Tokyo, 4259 Nagatsuta, Midori-ku, Yokohama 226-8501, Japan

[3]Institute for Open and Transdisciplinary Research Initiatives, Osaka University, 2-1 Yamada-oka, Suita, Osaka 565-0871, Japan

[4]Research Organization of Science and Technology, Ritsumeikan University, 1-1-1 Noji-higashi, Kusatsu, Shiga 525-8577, Japan

[5]Materials Integration Laboratories, AGC Inc., 1-1 Suehiro-cho, Tsurumi-ku, Yokohama 230-0045, Japan

[6]Graduate School of Arts and Sciences, The University of Tokyo, 3-8-1 Komaba, Meguro-ku, Tokyo 153-8902, Japan

[7]Center for Materials Research by Information Integration (CMI2), Research and Services Division of Materials Data and Integrated System (MaDIS), National Institute for Materials Science (NIMS), 1-1 Namiki, Tsukuba, Ibaraki 305-0044, Japan

[8]Department of Physical Sciences, Ritsumeikan University, 1-1-1 Noji-higashi, Kusatsu, Shiga 525-8577, Japan

[9]Graduate School of Engineering, Kyoto University, Kyotodaigaku-Katsura, Nishikyo-ku, Kyoto 615-8520, Japan

*Corresponding author
Email: mori@ims.tuskuba.ac.jp (T. Mori)
†These authors contributed equally to this work.

**Abstract**

Boson peak (BP) dynamics refers to the universal excitation in the terahertz region of glass. In this study, the universal dynamics of BP were quantitatively evaluated in various glassy materials based on the heterogeneous elasticity theory (HET), and the determinants of BP were successfully extracted. A strong correlation was observed between the maximum possible coarse-graining wavenumber, which is a determinant of the BP in the HET, and the first sharp diffraction peak (FSDP) wavenumber, which is a characteristic index of the medium-range order in glasses. The results indicate that the behaviour of BP in glass can be quantitatively understood in the following two steps. First, the FSDP representing the largest structural correlation in glass is dominantly used to determine the unit size of the elastic modulus heterogeneity, and second, the magnitude of the elastic modulus fluctuation is used to determine the frequency and intensity of the BP.

## Introduction

A universal excitation called the boson peak (BP) is observed in glass-forming materials in the terahertz (THz) region, which is the frequency band at the end of sound waves [1,2]. The BP is observed as the excess vibrational density of states (v-DOS, $g(\omega)$) deviating from the Debye model that describes crystal sound waves. Because the v-DOS of the Debye model is proportional to $\omega^{D-1}$ for $D$-dimensional materials, where $\omega$ is the angular frequency, the BP appears as a peak in the $g(\omega)/\omega^{D-1}$ spectrum.

BP is the origin or trigger of various phenomena related to glassy materials, such as their ultralow thermal conductivity compared to crystals [3], microscopic plastic deformation of glass [4] and the beginning of THz band optical absorption in glass [5,6]. Thus, BP excitation significantly contributes to the thermal, mechanical, and optical properties of glass, and the understanding of its mechanism and its applications are expected to be paramount. However, despite numerous experiments [3,5–13] including computer simulations [4,14–19], and theoretical proposals [20–25] on the origin of BP over the last few decades, no consensus has been reached.

As a BP theory, heterogeneous elasticity theory (HET) is well known and allows for a quantitative interpretation of BP behaviour [20,26–28]. The HET model incorporates spatial fluctuations into the shear modulus in the equation of motion of an isotropic elastic body, i.e. the 'Navier equation', considering the structural disorder of glass. The basic equations are as follows:

$$\frac{\partial^2 \boldsymbol{u}(\boldsymbol{r},t)}{\partial t^2} = \frac{1}{\rho}\boldsymbol{\nabla}\left(\left(K + \frac{4}{3}G(\boldsymbol{r})\right)\boldsymbol{\nabla}\cdot\boldsymbol{u}(\boldsymbol{r},t)\right) - \frac{1}{\rho}\boldsymbol{\nabla}\times\left(G(\boldsymbol{r})\boldsymbol{\nabla}\times\boldsymbol{u}(\boldsymbol{r},t)\right), \qquad (1)$$

where $\boldsymbol{u}(\boldsymbol{r},t)$ is the positional $\boldsymbol{r}$ and time $t$ dependent displacement in the continuum. $K$ and $G(\boldsymbol{r})$ are the bulk modulus and the shear modulus with spatial fluctuations, respectively, and $\rho$ is the mass density. In Navier or simple wave equations, the elastic moduli ($K$ and $G$) and density $\rho$ are treated as constants. This is equivalent to treating a homogeneous continuum or crystal. However, the HET model incorporates spatial heterogeneity in $G$ owing to the structural disorder of the glassy system.

A technique used to solve Eq. (1) is the coherent potential approximation (CPA), which is a mean-field approximation [26,28,29]. This approximation makes it possible to replace the spatially heterogeneous shear modulus ($G(r)$ in Eq. (1) or $\tilde{G}_i$ in Eq. (2) (see Methods)) with a spatially uniform frequency-dependent complex shear modulus $G(\omega)$. The CPA model can treat large heterogeneities in the elastic modulus and can therefore reproduce the large BP intensity of some glasses (e.g. silica glass), which could not be reproduced in previously reported HET models (see Figure 1 of ref. 8). In the CPA model [26,28], two main determinants of BP are present: 1) the magnitude of the elastic modulus fluctuations, and 2) the length of the spatial correlation of the elastic heterogeneity. However, understanding how these factors are related to the actual physical properties of glass remains open.

Furthermore, connecting the parameters of the model with real physical properties is required for the HET to explain the BP of real materials. In particular, the relationship between the elastic heterogeneity and the microscopic glass structure, especially the medium-range order of the glass, is not understood. The first sharp diffraction peak (FSDP) is a representative indicator of medium-range order. The FSDP appears as a strong peak at the lowest angle in the structure factor $S(k)$ of the glass [30,31], where $k$ is the wavenumber. The reciprocal of FSDP indicates the size of the pseudo-lattice of the glass, corresponding to the size of the unit cell in the case of a crystal. Although the relationship between the FSDP and BP has been discussed for various glasses [32–35], a universal relationship is yet to be demonstrated. Therefore, even in the HET framework, the relationship between the descriptors of the theory and the medium-range order has not been established. This obscures the picture of HET in actual glass, and in some cases, causes HET to be rejected [36,37].

In this study, we first performed CPA analysis based on HET on typical glasses, i.e. inorganic silica glass ($SiO_2$) and organic glycerol glass, to extract the nanomechanical properties that can reproduce the BP of glasses. Subsequently, we performed CPA on various glasses to investigate the relationship between the medium-range order of glasses and the spatial correlation length of elastic heterogeneity, which is a determinant of BP in the CPA model. We found that the spatial correlation of the elastic modulus was dominated by the pseudo-lattice size, which is determined by the FSDP of the glass. Finally, we concluded that the two properties of BP—frequency and intensity—can be explained by two factors, namely the pseudo-lattice size of the glass and the elastic heterogeneity magnitude.

## Methods

### Coherent potential approximation (CPA) analysis

To obtain the normalised effective shear modulus $\tilde{G}(z)$ ($z = \omega + i\epsilon$), the following CPA equation was solved numerically [28]:

$$\left\langle \frac{\tilde{G}_i - \tilde{G}(z)}{1 + \frac{1}{3}[\tilde{G}_i - \tilde{G}(z)]\Lambda(k_e, z)} \right\rangle_P = 0, \tag{2}$$

where $\tilde{G}_i$ is the mass-density-normalised shear modulus with spatial fluctuations defined by $\tilde{G}_i = G_i/\rho \equiv G(\mathbf{r}_i)/\rho$. $\tilde{G}(z)$ is the mass-density-normalised effective shear modulus, defined as $\tilde{G}(z) = G(z)/\rho$. In Eq. (2), to reproduce the BP frequency and intensity of various glasses, we used the following log-normal distribution function:

$$P(\tilde{G}_i, \tilde{G}_0, \sigma) = \frac{1}{\sigma\sqrt{2\pi}} \frac{1}{\tilde{G}_i} \exp\left\{-\frac{1}{2\sigma^2}[\ln(\tilde{G}_i/\tilde{G}_0)]^2\right\}, \tag{3}$$

where $\tilde{G}_0 = G_0/\rho$ is the mass-density-normalised geometric mean of $G_i$. $\sigma^2$ is the disorder parameter of spatial distribution of shear modulus, which is related to variance $\mathrm{Var}[\tilde{G}_i]$ by $\sigma^2 = \ln\{1 + \mathrm{Var}[\tilde{G}_i]/\langle \tilde{G}_i \rangle_P^2\}$ [28]. Although we recognise that the distribution of elastic heterogeneity in real glasses does not generally follow a log-normal distribution, we chose the log-normal distribution function to investigate the properties of BP (BP frequency and intensity) with two unified parameters ($k_e$ and $\sigma^2$) in the CPA. Notably, for small $\sigma^2$, the line shapes of the log-normal distribution and the Gaussian distribution show similar behaviour, as depicted in Supplementary Figure S1 and these distributions constitute a similar BP spectrum shape.

The integral kernel $\Lambda(k_e, z)$ in Eq. (2) is as follows:

$$\Lambda(k_e, z) = \frac{3}{k_e^3} \int_0^{k_e} dk\, k^4 \left(\frac{4}{3}\mathcal{G}_L(k, z) + 2\mathcal{G}_T(k, z)\right), \tag{4}$$

where $k_e$ is the maximum possible coarse-grained wavenumber. The reciprocal of $k_e$, i.e. the minimum possible coarse-graining length $\lambda_e$ has a physical interpretation as the correlation length of spatial fluctuations of the shear modulus [38,39]. The relationship between $k_e$ and $\lambda_e$ is as follows [28]:

$$k_e = \sqrt[3]{2\pi^2}/\lambda_e. \tag{5}$$

Green's functions for the longitudinal wave $\mathcal{G}_L(k, \omega)$ and transverse wave $\mathcal{G}_T(k, \omega)$ are defined as follows:

$$\mathcal{G}_L(k, z) = \frac{1}{-z^2 + k^2 v_L(z)^2}, \tag{6a}$$

$$\mathcal{G}_T(k, z) = \frac{1}{-z^2 + k^2 v_T(z)^2}, \tag{6b}$$

where the complex frequency-dependent transverse sound velocity $v_T(z)$ and complex frequency-dependent longitudinal sound velocity $v_L(z)$ are related to $\tilde{G}(z)$ and the experimental bulk modulus $K_{\exp}$ by the following relationships:

$$v_T(z)^2 = \tilde{G}(z), \tag{7a}$$

$$v_{\text{L}}(z)^2 = \frac{K_{\text{exp}}}{\rho} + \frac{4}{3}\tilde{G}(z). \tag{7b}$$

The input parameters for solving Eq. (2) show the experimental data of transverse $v_{\text{T}}$ and longitudinal $v_{\text{L}}$ sound velocities. When $z = 0$ and $v_{\text{T}}^2 = \tilde{G}(0)$, the relationship between $\tilde{G}(0)/\tilde{G}_0$ and $\sigma$ as an implicit function is detailed in the Appendix of Ref. 28.

The vibrational density of states $g(\omega)$ is expressed as follows:

$$g(\omega) = \text{Im}\left[\frac{2\omega}{3\pi}\frac{3}{k_{\text{D}}^3}\int_0^{k_{\text{D}}} dk\, k^2\big(\mathcal{G}_{\text{L}}(k,\omega) + 2\mathcal{G}_{\text{T}}(k,\omega)\big)\right], \tag{8}$$

where the Debye wavenumber $k_{\text{D}}$ is defined as $k_{\text{D}} = \sqrt[3]{6\pi^2 N/V}$. $N$ and $V$ are the number of particles and the volume of the system, respectively.

The fits for SiO$_2$ and glycerol were performed using the least squares method (Figure S1). For other glasses, $k_{\text{e}}$ and $\sigma$ were determined to reproduce the BP frequency and intensity (see Table S1).

## Results

As shown in the BP spectra $g(\omega)/\omega^2$ in Figure 1(a), the BP frequencies of typical inorganic SiO$_2$ and typical organic glycerol glass are both located at approximately 1 THz, and their BP heights are almost the same [40,41]. However, the Debye levels of the two materials are significantly different, and the Debye level of SiO$_2$ is low because of its large sound velocity. This indicates that the BP intensity of SiO$_2$ is much higher than that of glycerol. This difference is clarified in the normalised BP spectrum shown in Figure 1(b). The horizontal and vertical axes were normalised using the Debye frequency $\omega_{\text{D}} = [18\pi^2(N/V)/(2v_{\text{T}}^{-3} + v_{\text{L}}^{-3})]^{1/3}$ and Debye level $A_{\text{D}} = 3/\omega_{\text{D}}^3$, respectively. $N$ and $V$ are the number of particles and volume of the system, respectively. Moreover, $v_{\text{T}}$ and $v_{\text{L}}$ are the transverse and longitudinal sound velocities, respectively. The normalised spectrum shows that SiO$_2$ has a lower BP frequency and a much higher BP intensity than glycerol.

To reproduce these spectra using HET and discuss the differences in the properties of BP using unified parameters, we fitted the experimental spectra with the CPA equation using the log-normal distribution function with the maximum possible coarse-graining wavenumber $k_{\text{e}}$ and the disorder parameter $\sigma^2$ as independent parameters (see Methods). The $G_0$, which is a geometric mean of $G(r)$, in the log-normal distribution function is implicitly related to $\sigma$ [28]. The extracted parameters $G_0$, $k_{\text{e}}$ and $\sigma$ are summarised in Table 1, and the obtained complex moduli $G(\omega)$ of SiO$_2$ and glycerol are shown in Supplementary Figure S1.

The $g(\omega)$ was calculated using Green's function, including $G(\omega)$ (see Eq. (8) in Methods), and the obtained $g(\omega)/\omega^2$ is shown in Figure 1(b) using solid lines. The BP frequency and intensity of the calculated spectra were both in good agreement with the experimental results.

Regarding the line shapes of CPA, although the line shape of glycerol agrees well with the experimental spectrum, the result of $SiO_2$ considerably deviates from the experimental result. Modifications are required in the CPA to correct this deviation, such as applying a probability density function that reproduces the v-DOS of silicate glasses; however, this is beyond the scope of this study.

To understand the origin of the different BP behaviours of $SiO_2$ and glycerol, we discuss the $k_e - \sigma^2$ dependence of the normalised BP frequency and intensity, as shown in Figure 2. The normalised $k_e$ values of the two substances were found to be almost the same; however, in contrast, $\sigma^2$ of $SiO_2$ is 2.3 times larger than that of glycerol. The normalised BP frequency and intensity obtained by solving the CPA equation with $v_L/v_T = 2$, which is a typical ratio for various glasses, as input parameters are shown in the colour map. This colour map shows that when $k_e$ is fixed and $\sigma^2$ increases, the BP frequency decreases (Figure 2(a)), whereas the BP intensity increases (Figure 2(b)). Thus, comparing the two materials, a large $\sigma^2$ of $SiO_2$ decreases the BP frequency and increases the BP intensity compared with glycerol.

Furthermore, we visualised the obtained parameters $k_e$ and $\sigma$ using the glass structure (see SI) in real space and discussed their physical meaning. First, in Figure 3, the reciprocal of the maximum possible coarse-graining wavenumber $k_e$, i.e. the minimum possible coarse-graining wavelength $\lambda_e$ defined by Eq. (5), is displayed on an actual scale as the lattice size of the grid lines. The lattice sizes indicate $\lambda_e$ of each material, which is ~3 Å for both materials. In addition, the dimensions of the glass structures in Figure 3 indicate wavelengths of $\lambda_{BP-SiO2} = 38.0$ Å and $\lambda_{BP-Gly} = 19.6$ Å of the BP modes of $SiO_2$ and glycerol, respectively. Here, the wavelength of the BP mode $\lambda_{BP}$ is the same as what is generally called the characteristic length of BP. The wavelength is defined by $\lambda_{BP} = 2\pi v_T/\omega_{BP}$.

Moreover, we found that the $\lambda_{BP}$ values of $SiO_2$ and glycerol were considerably different, and the ratio of $\lambda_{BP}$ to $\lambda_e$ was as large as 12 for $SiO_2$ and 6 for glycerol. This difference in the $\lambda_{BP}/\lambda_e$ ratio is due to the difference in the magnitude of the fluctuation of the shear modulus of the two materials, i.e. the difference in the magnitude of the disorder parameter $\sigma^2$. This is shown in the glass structures depicted in Supplementary Figure S3 as colour distributions based on the log-normal distribution function obtained from the CPA analysis of each material. The greater the magnitude of the fluctuations in the shear modulus, the longer the wavelength and lower the normalised frequency of the BP mode, in which a rapid increase in the imaginary part of $G(\omega)$ starts toward the higher frequency in the effective medium (Supplementary Figures S2(a) and S2(c)). Consequently, a large $\sigma^2$ increases $\lambda_{BP}/\lambda_e$, i.e. it decreases the normalised BP frequency.

Next, we considered the physical meanings and origins of $k_e$ and its reciprocal $\lambda_e$. Previous research [43] has shown that when shear modulus fluctuations have a spatial correlation, the reciprocal of $k_e$ approximately represents the length of the spatial correlation of the shear modulus. Considering this, the FSDP in the $S(k)$, which is an indicator of the medium-range order of glasses, is a natural candidate for the origin of the minimum length of the spatial correlation of the elastic modulus in glasses. Supplementary Figure S4 shows the structural factors $S(k)$ of SiO$_2$ and glycerol. As summarised in Table 1, the FSDP wavenumbers $k_{\text{FSDP}}$ (or reciprocal $\lambda_{\text{FSDP}} = 2\pi/k_{\text{FSDP}}$) of SiO$_2$ and glycerol appear to agree well with their respective $k_e$ (or $\lambda_e$).

The generality of this relationship was verified by performing CPA on various glasses and examining the correlation between $k_e$ and $k_{\text{FSDP}}$. As shown in Figure 4, a strong positive correlation was observed between $k_{\text{FSDP}}$ and $k_e$ in various organic-to-inorganic glasses. Furthermore, the values of $k_e$ and $k_{\text{FSDP}}$ are of the same order. Note that we do not claim all glasses fall exactly on the fitting curve but emphasise that the fitting curve is indicative of the overall trend. These results lead to the following conjecture: 'The minimum possible coarse-graining size of the elastic modulus of glass is dominantly determined by FSDP: The BP is caused by the inhomogeneous elasticity fluctuating for each size of the pseudo-lattice formed by FSDP'.

The relationship between the medium-range order in glasses and BP has been discussed [32–35]. However, the correlation between $k_{\text{FSDP}}$ and $k_{\text{BP}} = \omega_{\text{BP}}/v_{\text{T}}$ is not simple. Moreover, they appear to be uncorrelated, as shown in Supplementary Figure S5. However, based on our hypothesis, the relationship between $k_{\text{FSDP}}$ and $k_{\text{BP}}$ can be described in the following two steps. First, the minimum possible coarse-graining length $\lambda_e$ of the elastic modulus is dominantly determined by $\lambda_{\text{FSDP}}$. As explained by the HET framework and shown in Figure 5, the ratio $k_{\text{BP}}/k_e$ decreases as the disorder parameter $\sigma^2$ increases. In other words, as mentioned above, by increasing $\sigma^2$, $\lambda_{\text{BP}}/\lambda_e$ increases or the normalised BP frequency decreases. In previous studies [32–35], the parameters $k_e$ and $\sigma^2$ could not be separated; however, we clarified the quantitative relationship between $k_e$, a determinant of BP in the CPA, and the FSDP.

Furthermore, the scenario obtained in this study is applicable to the BP behaviour of simple glass with a Lennard–Jones (LJ) potential obtained by molecular dynamics (MD) simulations. We analysed the LJ glass [44] using the CPA equation, and the result is plotted by an open circle in Figure 4. In the $S(k)$ of the LJ glass, only one peak is present, which indicates the correlation between the neighbouring two atoms of simple glass (see Supplementary Figure S4). Therefore, we regarded the $k$ indicating the position of one peak as $k_{\text{FSDP}}$. The result of the LJ glass shown in Figure 4 deviated from the glasses with FSDP, and the ratio of $k_{\text{FSDP}}/k_e$ is 1.81 with $k_e/k_D = 0.238$. Furthermore, $\lambda_{\text{BP}}/\lambda_e = 5.5$ (see Supplementary Figure S6 and Supplementary Table S1). A recent study of MD simulation for simple glasses by Hu and Tanaka [36,37] suggested that the HET scenario does not hold because the spatial correlation size of the spring-constant heterogeneity in LJ glasses is mismatched with the wavelength of the string-like structure of the BP mode. The resolution of the spring-constant heterogeneity in their results (the resolution of the yellow-coloured area in Figure 5 of Ref. 36) is approximately two atoms, and the wavelength of BP is approximately 15 atoms. However, our analysis shows it is exactly the ratio of $\lambda_{\text{BP}}/\lambda_{\text{e}}$ as shown in Supplementary Figure S6, and the results of simple glasses can be also explained by the HET scenario.

In addition to the LJ glass, we performed CPA analysis on a physical gel system in MD simulations [44]. The physical gel system utilised the LJ potential, but its density is anomalously lower than that of typical LJ glass (see Supplementary Figure S7 and figure 1(d) of ref. 42). Regarding structure, the system has two peaks in $S(k)$ (see Supplementary Figure S4). The peak for the higher wavenumber represents the correlation between the neighbouring two atoms, also existing in LJ glass. The other peak for the lower wavenumber appears due to voids that are unique to physical gels as shown in Supplementary Figure S7. The results of CPA analysis for the physical gels assuming $k_{\text{FSDP}}$ as the lower peak are shown in Figure 4 by open squares. The characteristics of the physical gel results are located below the fitting curve of FSDP glasses. Consequently, physical gels, FSDP glasses and LJ glasses are in different regions of the $k_{\text{e}}/k_{\text{D}} - k_{\text{FSDP}}/k_{\text{D}}$ dependence. This is presumably due to the different properties of the lowest peaks in $S(k)$ for the three types of amorphous materials. It may seem unusual that the characteristic length of the heterogeneity of the elastic modulus has a strong relationship with the structure factor, i.e., density fluctuations. However, as shown in the CPA equation (Eq. (2)), HET is fundamentally a theory that addresses the inhomogeneity of the velocity field. Furthermore, the transverse sound velocity fluctuation $v_{\text{T}}(r)$ is related to the shear modulus $G(r)$ and mass density $\rho(r)$ fluctuations by $v_{\text{T}}(r)^2 = G(r)/\rho(r)$, and the mass density heterogeneity is essentially included in the velocity fluctuations. In the case of physical gels, the BP intensity is quite small (see Fig. 9 of ref. 42) even though the heterogeneity of the elastic modulus is noticeably large due to the presence of voids. The low BP intensity of physical gel is likely attributed to the small fluctuations of the average velocity field, as the elastic modulus and mass density are proportionally lower in the areas where voids are present.

In this study, we could not identify the physical origin of $\sigma$. Qualitatively, however, the presence of voids in the glass or the lower coordination number would produce a small local shear modulus and increase $\sigma$.

## Discussion

In this study, the determinants of BP in the CPA based on the HET were extracted by quantitatively reproducing the typical BP behaviour of glasses (BP frequency and intensity). Regarding the physical origin of the maximum possible coarse-graining wavenumber $k_{\text{e}}$, which is a determining factor of BP, we found that $k_{\text{e}}$ and $k_{\text{FSDP}}$, which is a characteristic index of the medium-range order of glasses, showed not only a strong positive correlation, but also comparable values. This result implies that the unit of modulus heterogeneity is dominated by FSDP or the lowest peak in $S(k)$, exhibiting a pseudo-lattice-like structural correlation in the glass. Furthermore, the BP frequency and intensity can be understood quantitatively based on the magnitude of the elastic modulus fluctuation for that unit.

Although a correlation between $k_\mathrm{e}$ and FSDP was observed, quantitative elucidation of the physical origin of $\sigma$, which is qualitatively attributed to local connectivity, is left for future work. When the correlation between these determinants of BP in the HET and its physical properties is fully understood, a quantitative understanding of the physical phenomena caused by BP and the application of BP may be realised.

## Acknowledgements
We thank Zhiwen Pan, Lothar Wondraczek and Walter Schirmacher for their useful discussions. We would like to thank Shogo Bando for his assistance with the CPA analysis of LJ glass and for visualizing the results. This work was supported by JSPS KAKENHI Grant Nos. 23H01139 (to T.M.), 23H04495 (to H.M.) and 22K03543 (to H.M.); and JSPS Grant-in-Aid for Transformative-Research Areas (A) 'Hyper-Ordered Structures Science' Grant Nos. 20H05878 (to S. Kohara) and 20H05881 (to S. Kohara); and the AGC research collaboration (to T.M.); and GIC & NGF (to T.M.).

## Author contributions
T.M. conceived this study and wrote the paper, with contributions and discussions from all authors. D.K. and S.H.O performed CPA analysis for VDOS data. S. Kitani. and H. K. performed maximum entropy method analysis for low-temperature specific heat data. S. Kohara. performed reverse Monte Carlo simulations to obtain the SiO$_2$ glass structure. T.M., H.M., H.H., Y.F., A.K., A.M., Y.Y., and S. Kojima evaluated the determinant factors of BP. All authors were involved in manuscript revisions.

## Conflict of interest
The authors declare no competing interests.

## Data availability
The data that support the findings of this study are available from the corresponding author upon reasonable request.

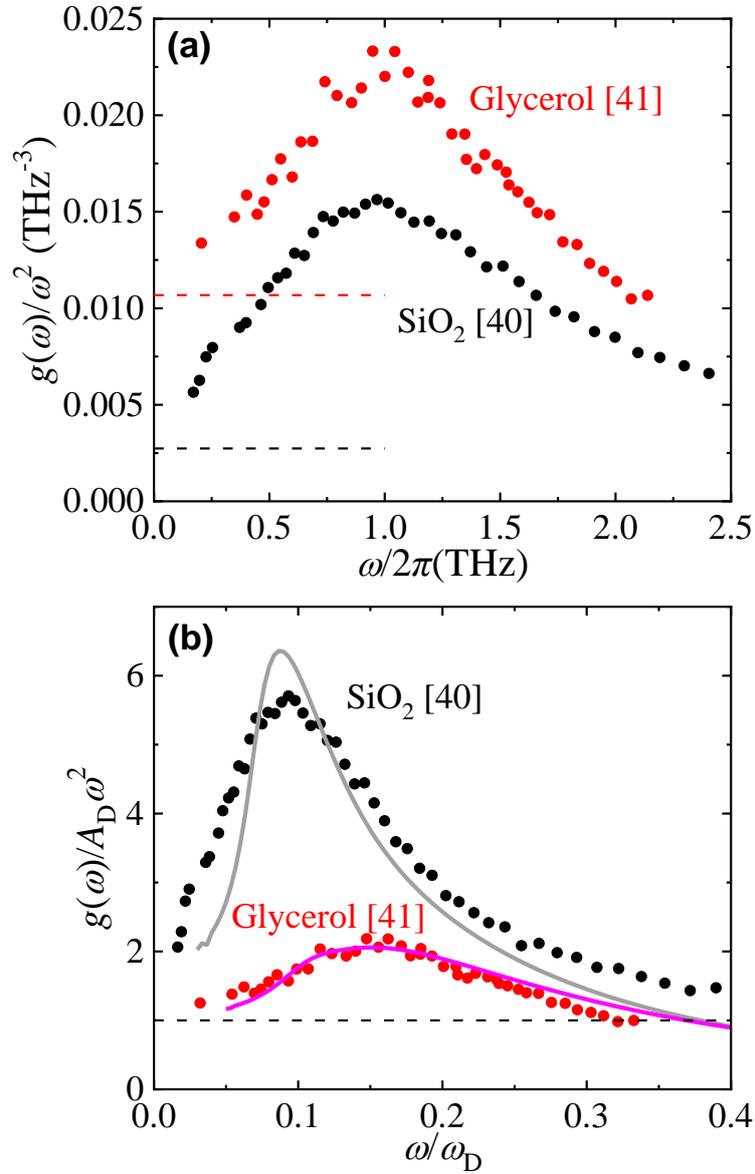

**Figure 1 | Boson peak spectra of SiO$_2$ and glycerol. a** Boson peaks in the $g(\omega)/\omega^2$ spectrum of SiO$_2$ and glycerol. The data of SiO$_2$ and glycerol obtained by inelastic neutron scattering measurements are quoted from the literature [40,41], respectively. Black and red dashed lines indicate Debye levels of SiO$_2$ and glycerol, respectively. **b** Normalised BP plots of SiO$_2$ and glycerol. The vertical axis was normalised to the Debye level, and the horizontal axis was normalised to the Debye frequency. The grey and pink solid lines show the CPA results for SiO$_2$ and glycerol, respectively.

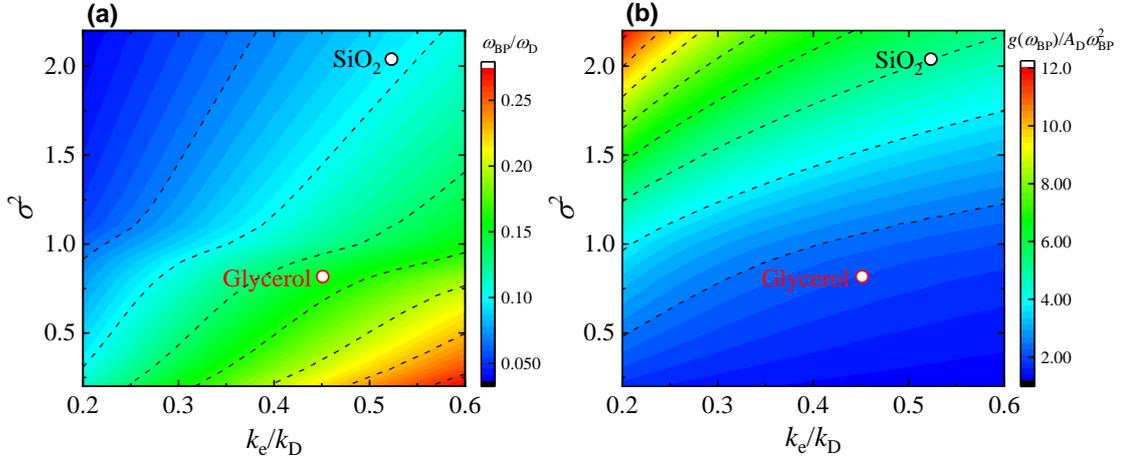

**Figure 2 | $k_e - \sigma^2$ dependence of normalised BP frequency and normalised BP intensity.** The locations of **a** normalised BP frequency and **b** normalised BP intensity of SiO$_2$ (black open circles) and glycerol (red open circles) on the $k_e - \sigma^2$ planes obtained by CPA analysis. The input parameters for each glass used for the CPA analysis are summarised in Table S1. Colour maps indicate the $k_e - \sigma^2$ dependence of the normalised BP frequency and normalised BP intensity obtained by CPA analysis for Eq. (2) with $v_L/v_T = 2$, which is a typical ratio for various glasses. The resolutions of $k_e$ and $\sigma^2$ are 0.01 and 0.1, respectively. Contours are drawn by dashed lines for the normalised BP frequency every 0.032 and for the normalised BP intensity every 1.30.

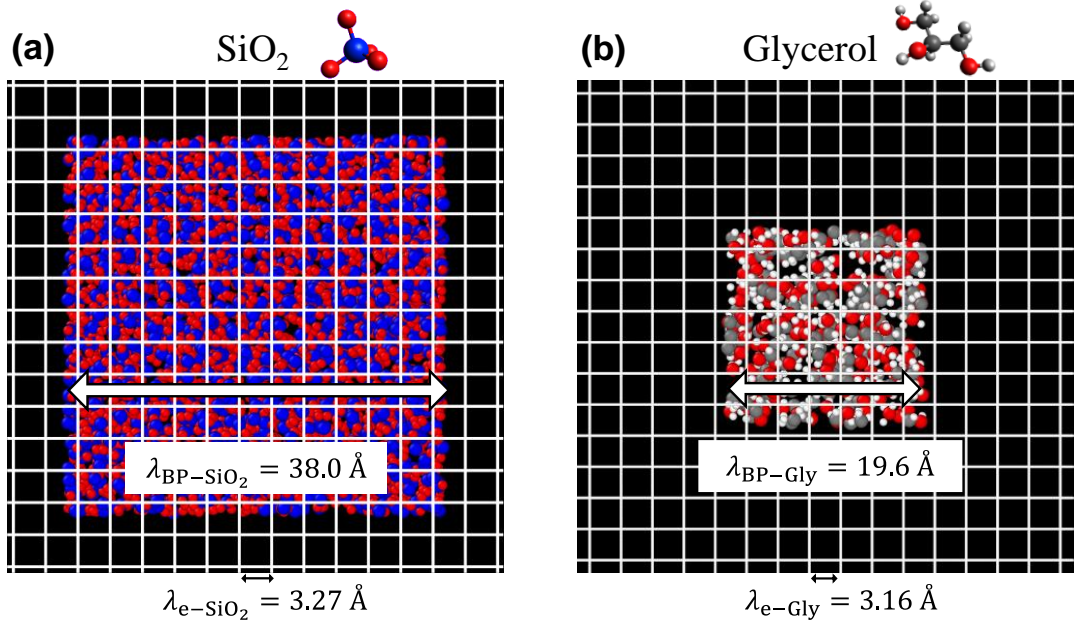

**Figure 3 | BP mode wavelength $\lambda_{BP}$ and minimum possible coarse-graining length $\lambda_e$ of SiO₂ and glycerol.** Glass structures of **a** $SiO_2$ and **b** glycerol. The dimension of each glass structure indicates the wavelength of each BP mode. The lattice size of the grid lines indicates the minimum possible coarse-graining length $\lambda_e$. The visualization of glass structures was carried out through Ovito [42].

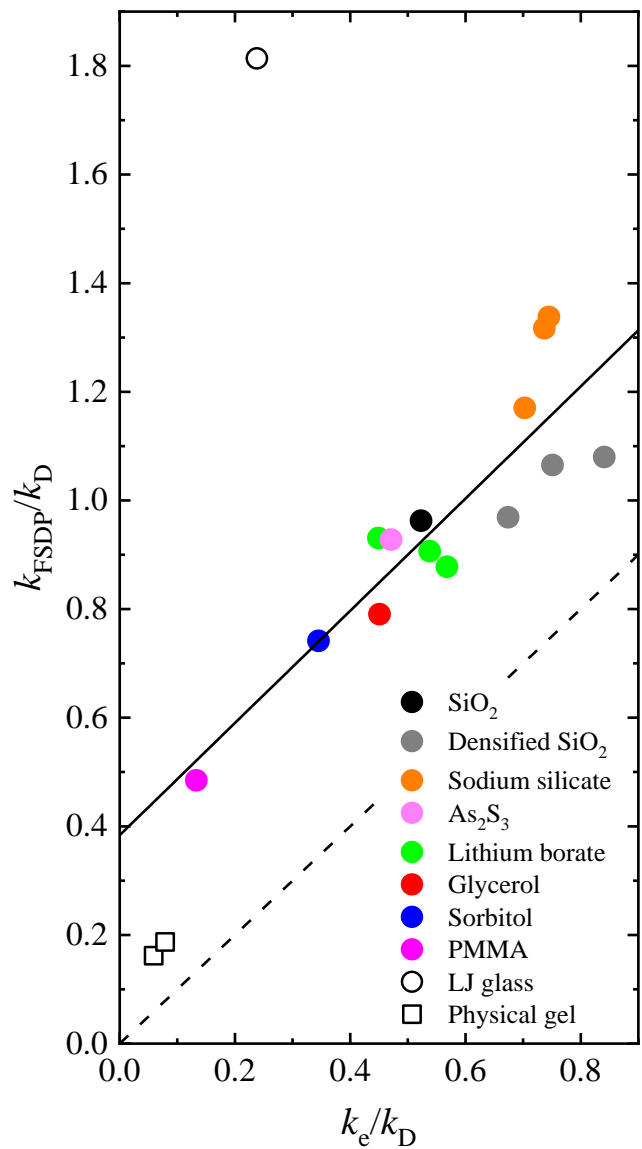

**Figure 4 | Relationship between the FSDP wavenumber $k_{FSDP}$ and the maximum possible coarse-graining wavenumber $k_e$ for various glasses.** Both $k_{FSDP}$ and $k_e$ were normalised by the Debye wavenumber $k_D$. The parameters for each glass are summarised in Table S1. The solid line indicates the fitting function obtained by the least-squares method for FSDP glasses, where $k_{FSDP}/k_D = 1.0 \pm 0.16\ k_e/k_D + 0.38 \pm 0.094$, and the correlation coefficient is 0.883. The dashed line indicates a proportional line with slope 1. Open circles and open squares show the results of LJ glass and physical gels, respectively.

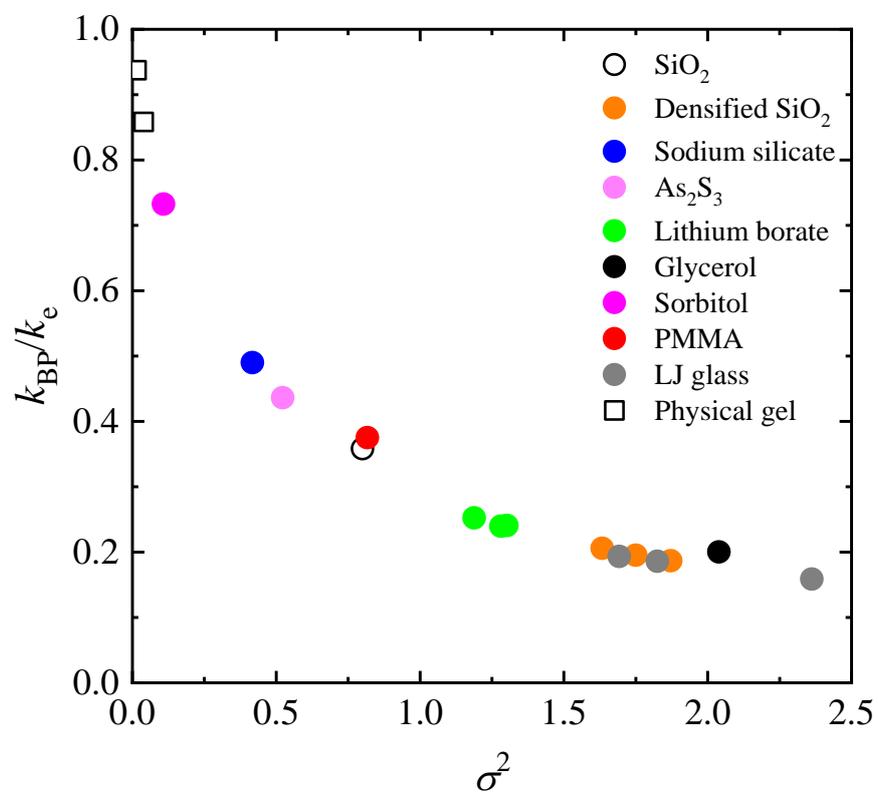

**Figure 5 | Relationship between the BP wavenumber $k_{BP}$ and the disorder parameter $\sigma^2$ for various glasses.** $k_{BP}$ was normalised by the maximum coarse-grained wavenumber $k_e$. The parameters for each glass are summarised in Supplementary Table S1.

**Table 1 | Extracted parameters in CPA analysis of the SiO₂ and glycerol.** Normalised maximum possible coarse-graining wavenumber $k_e/k_D$, minimum possible coarse-graining wavelength $\lambda_e$, disorder parameter of spatial distribution of shear modulus $\sigma^2$, geometric mean of spatial distribution of shear modulus $G_0$, ratio of $k_{BP}$ to $k_e$, BP wavelength $\lambda_{BP}$, normalised FSDP wavenumber $k_{FSDP}/k_D$, and Debye wavenumber $k_D$.

|         | $k_e/k_D$ | $\lambda_e$ (Å) | $\sigma^2$ | $G_0$ (GPa) | $k_{BP}/k_e$ | $\lambda_{BP}$ (Å) | $k_{FSDP}/k_D$ | $k_D$ (Å⁻¹) |
|---------|-----------|-----------------|------------|-------------|--------------|---------------------|-----------------|--------------|
| SiO₂    | 0.523     | 3.27            | 2.04       | 57.2        | 0.200        | 38.0                | 0.96            | 1.58[1]      |
| Glycerol| 0.451     | 3.16            | 0.82       | 5.43        | 0.375        | 19.6                | 0.79[2]         | 1.90[3]      |

[1]Reference 31.
[2]Reference 45.
[3]Reference 46.

# Supplementary Information

# Relationship between the boson peak and first sharp diffraction peak in glasses

Dan Kyotani[1,†], Soo Han Oh[1,†], Suguru Kitani[2], Yasuhiro Fujii[3,4], Hiroyuki Hijiya[5], Hideyuki Mizuno[6], Shinji Kohara[7], Akitoshi Koreeda[8], Atsunobu Masuno[9], Hitoshi Kawaji[2], Seiji Kojima[1], Yohei Yamamoto[1], Tatsuya Mori[1*]

[1]Department of Materials Science, University of Tsukuba, 1-1-1 Tennodai, Tsukuba, Ibaraki 305-8573, Japan

[2]Materials and Structures Laboratory, Institute of Integrated Research, Institute of Science Tokyo, 4259 Nagatsuta, Midori-ku, Yokohama 226-8501, Japan

[3]Institute for Open and Transdisciplinary Research Initiatives, Osaka University, 2-1 Yamada-oka, Suita, Osaka 565-0871, Japan

[4]Research Organization of Science and Technology, Ritsumeikan University, 1-1-1 Noji-higashi, Kusatsu, Shiga 525-8577, Japan

[5]Materials Integration Laboratories, AGC Inc., 1-1 Suehiro-cho, Tsurumi-ku, Yokohama 230-0045, Japan

[6]Graduate School of Arts and Sciences, The University of Tokyo, 3-8-1 Komaba, Meguro-ku, Tokyo 153-8902, Japan

[7]Center for Materials Research by Information Integration (CMI2), Research and Services Division of Materials Data and Integrated System (MaDIS), National Institute for Materials Science (NIMS), 1-1 Namiki, Tsukuba, Ibaraki 305-0044, Japan

[8]Department of Physical Sciences, Ritsumeikan University, 1-1-1 Noji-higashi, Kusatsu, Shiga 525-8577, Japan

[9]Graduate School of Engineering, Kyoto University, Kyotodaigaku-Katsura, Nishikyo-ku, Kyoto 615-8520, Japan

*Corresponding author
Email: mori@ims.tuskuba.ac.jp (T. Mori)
†These authors contributed equally to this work.

## Content

This file contains Supplementary Figures S1–S7, Supplementary Table S1, and Supplementary References 1–23.

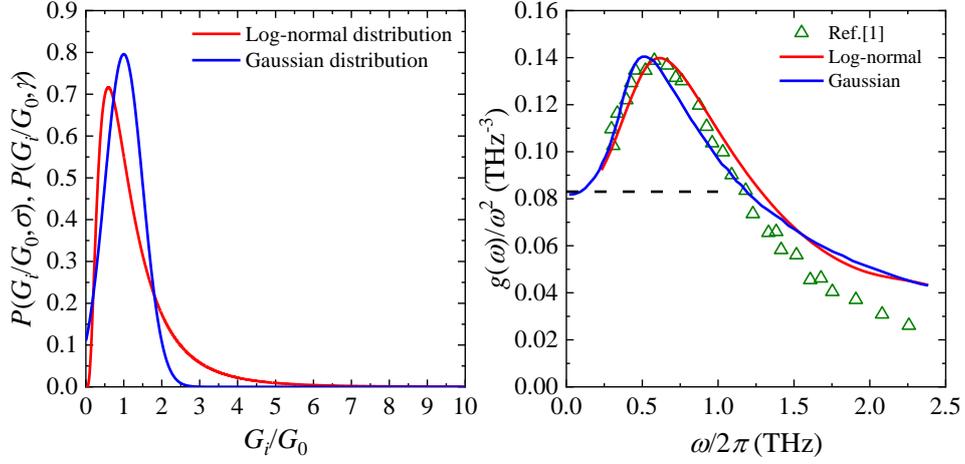

**Figure S1. Comparison of BP spectra calculated using log-normal and normal distribution functions.** (Left) The log-normal (red line) and normal distribution (blue line) graphs of the probability density function of shear modulus used in HET to reproduce the BP spectrum of $As_2S_3$ glass. The parameters of the normal distribution function $P(G,\gamma) = \exp(-(G-G_0)^2/\gamma^2)/\sqrt{2\pi\gamma}$, where $\gamma = 0.352 \times v_T^4$ [km$^4$/s$^4$] with $v_T = 1.4$ [km/s] and $G_0 = 7.28$ GPa are from a previous study [1]. (Right) BP spectra $g(\omega)/\omega^2$ calculated using log-normal (red line) and normal distribution (blue line) functions for $As_2S_3$ glass. Black dashed line indicates Debye level of $As_2S_3$ glass. The spectrum from normal distribution (blue line) and experimental spectrum (green opened triangles) were extracted from a previous study [1].

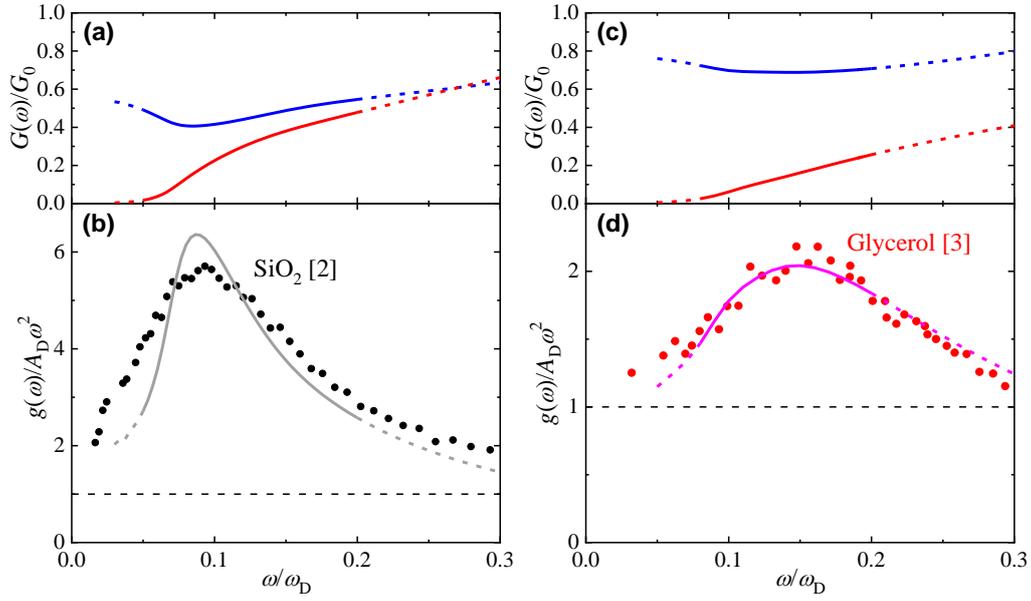

**Figure S2. CPA fit results for BP spectra of SiO₂ and glycerol.** Real (blue) and imaginary (red) part of $G(0)/G_0$ of **a** SiO$_2$ and **c** glycerol. The solid line part indicates the fitted range, and the rest is indicated by the dashed line. $G_0$ of SiO$_2$ and glycerol are 40.1 GPa and 6.4 GPa, respectively. Normalised BP plots of **b** SiO$_2$ and **d** glycerol. The vertical axis was normalised by Debye level, and the horizontal axis was normalised by Debye frequency. Grey and pink solid lines show the results of CPA analysis for SiO$_2$ and glycerol, respectively. The filled circles show the experimental data of SiO$_2$ (black) and glycerol (red), respectively. The data of SiO$_2$ and glycerol are quoted from previous studies [2, 3].

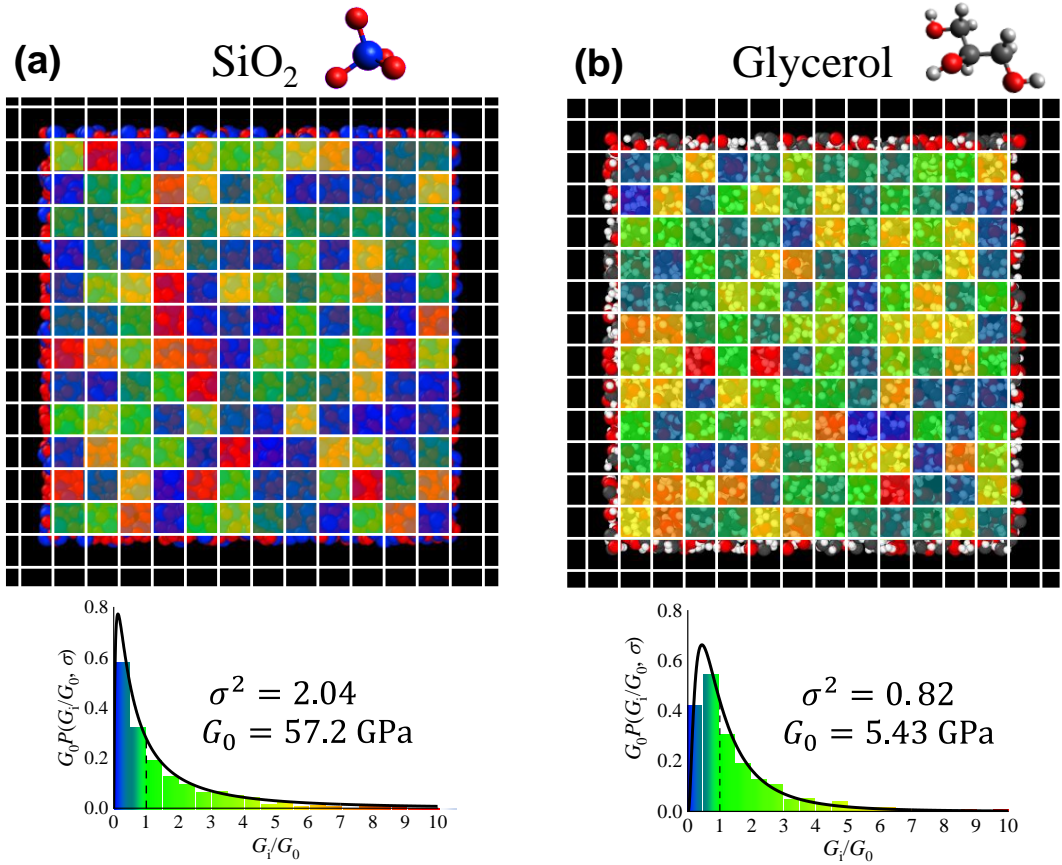

**Figure S3. Schematic of the spatial distribution of shear modulus $G_i$ of SiO$_2$ and glycerol.** Colour distributions at the upper part of the figure show the spatial distribution of $G_i/G_0$ following the log-normal distribution function on **a** SiO$_2$ and **b** glycerol structures. The lattice size of the grid lines indicates the minimum possible coarse-graining length $\lambda_\mathrm{e}$. The lower part of the figure shows the log-normal distribution functions of **a** SiO$_2$ and **b** glycerol obtained by CPA analysis. The colours used in the histogram correspond to the colours of the spatial distribution of $G_i/G_0$ shown in the upper part of the figure. The parameters of the lattice size $\lambda_\mathrm{e}$ and the log-normal distribution function ($G_0$ and $\sigma$) are summarised in Table 1. The structure of SiO$_2$ was obtained by reverse Monte Carlo simulation of the structure factors of SiO$_2$ [4]. The structure of glycerol was obtained by performing MD simulation of glycerol with the LAMMPS package [5] using Chelli's potential [6]. The visualisation of glass structures was conducted through OVITO [7].

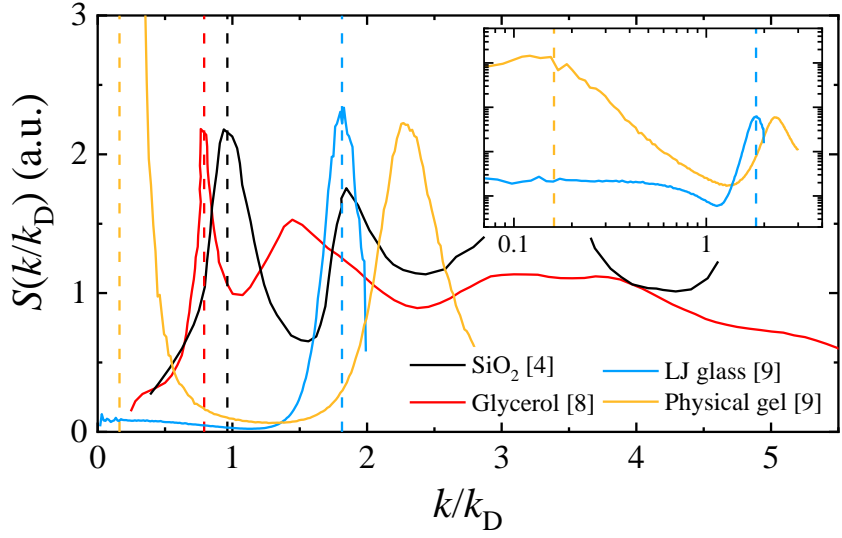

**Figure S4. Structural factors of SiO$_2$, glycerol, LJ glass and physical gel.** The horizontal axis of $k$ is normalised by $k_D$. The FSDPs of silica glass (black line) and glycerol (red line) are located at 0.96 and 0.76, respectively. The inset shows log-log plot of structure factors of LJ glass (cyan line) and physical gel (yellow line). The lowest peaks of LJ glass and physical gel with $\rho = 0.5$ are located at 1.81 and 0.16, respectively. The data of SiO$_2$ and glycerol were quoted from previous studies [4,8]. The data of LJ glass and physical gel were quoted from a previous study [9].

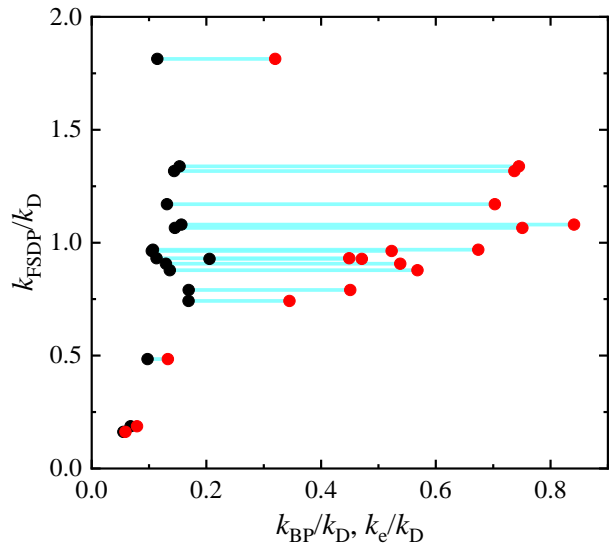

**Figure S5. Relationship between the FSDP wavenumber $k_{FSDP}$ and the BP wavenumber $k_{BP}$ for various glasses.** $k_{FSDP}$, $k_{BP}$ and $k_e$ were all normalised by the Debye wavenumber $k_D$. $k_{BP}/k_D$ (black circle) and $k_e/k_D$ (red circle) of the same substance are connected by a solid cyan line for visibility. The parameters for each glass are summarised in Supplementary Table S1.

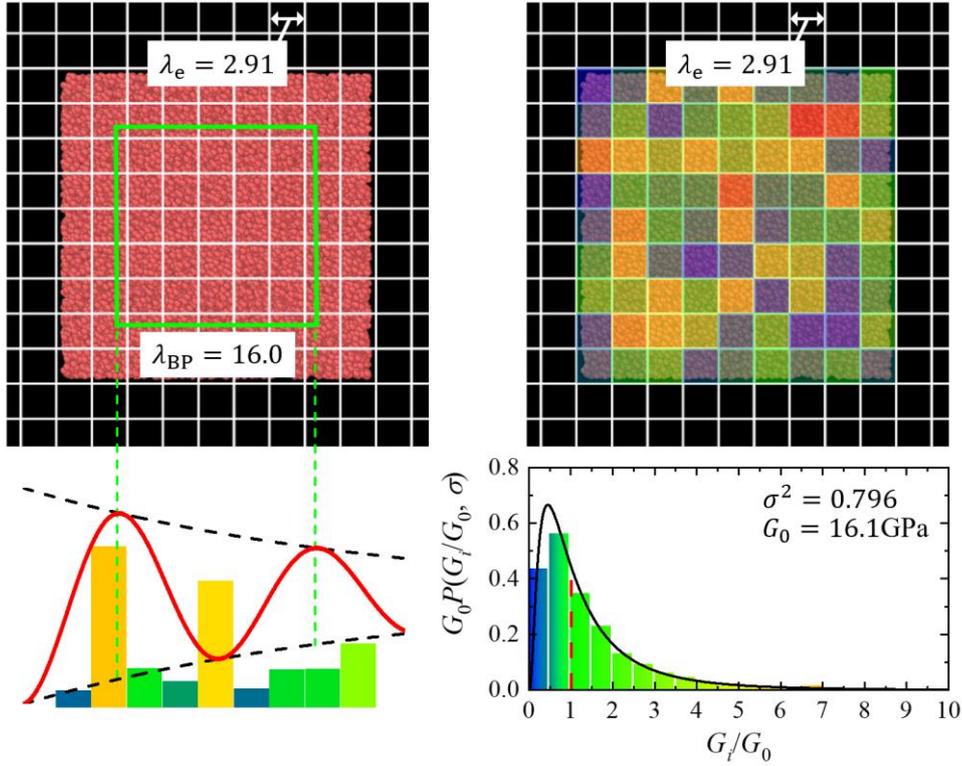

**Figure S6. Visualization of results of CPA analysis on LJ glass ($\rho$ = 1.0).** (Left upper) The lattice size of the grid lines and length of one side of the green square on the LJ glass structure indicate the minimum possible coarse-graining length $\lambda_e$ and BP wavelength $\lambda_{BP}$, respectively. (Left bottom) The red line represents a sound wave with a wavelength of $\lambda_{BP}$ that is attenuated by a random shear modulus distribution. (Right upper) The colour distribution shows the spatial distribution of $G_i/G_0$ following the log-normal distribution function on the LJ glass structure. The lattice size of the grid lines indicates the minimum possible coarse-graining length $\lambda_e$. (Right bottom) The log-normal distribution functions of LJ glass obtained by CPA analysis. The colours used in the histogram correspond to the colours of the spatial distribution of $G_i/G_0$ shown in the upper part of the figure. The parameters of the lattice size $\lambda_e$ and the log-normal distribution function ($G_0$ and $\sigma$) are summarised in Table 1.

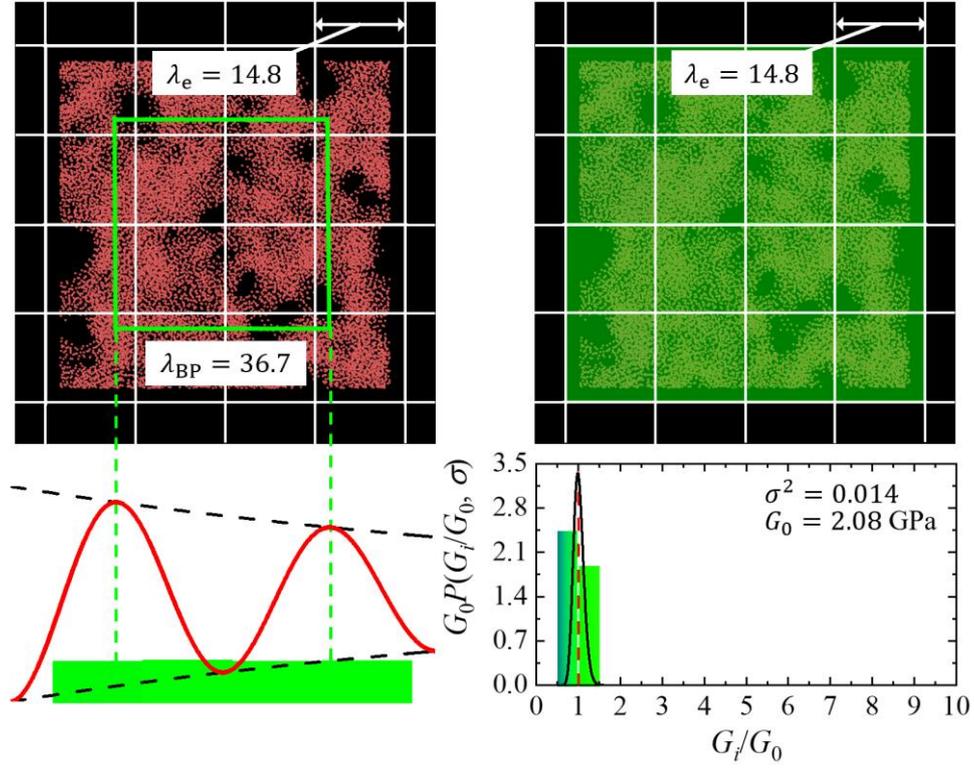

**Figure S7. Visualization of results of CPA analysis on physical gel ($\rho$ = 0.5).** (Left upper) The lattice size of the grid lines and length of one side of the green square on the physical gel structure indicate the minimum possible coarse-graining length $\lambda_e$ and BP wavelength $\lambda_{BP}$, respectively. (Left bottom) The red line represents a sound wave with a wavelength of $\lambda_{BP}$ that is attenuated by a random shear modulus distribution. (Right upper) Colour distribution shows the spatial distribution of $G_i/G_0$ following the log-normal distribution function on physical gel structure. The lattice size of the grid lines indicates the minimum possible coarse-graining length $\lambda_e$. (Right bottom) The log-normal distribution functions of LJ glass obtained by CPA analysis. The colours used in the histogram correspond to the colours of the spatial distribution of $G_i/G_0$ shown in the upper part of the figure. The parameters of the lattice size $\lambda_e$ and the log-normal distribution function ($G_0$ and $\sigma$) are summarised in Table 1.

**Table S1. Input and extracted parameters in the CPA analysis.** Mass density $\rho$, transverse $v_T$ and longitudinal $v_L$ sound velocities, normalised maximum possible coarse-graining wavenumber $k_e/k_D$, minimum possible coarse-graining wavelength $\lambda_e$, disorder parameter of spatial distribution of shear modulus $\sigma^2$, geometric mean of spatial distribution of shear modulus $G_0$, ratio of $k_{BP}$ to $k_e$, BP wavelength $\lambda_{BP}$, normalised FSDP wavenumber $k_{FSDP}/k_D$, and Debye wavenumber $k_D$. Sodium silicate glass, lithium borate glass and polymethyl methacrylate are abbreviated as NS, LiB and PMMA, respectively. $g(\omega)$ of densified $SiO_2$, NS, and LiB were determined from the specific heat data by maximum entropy method analysis [10].

| | $\rho$ (g/cm$^{-3}$) | $v_T$ (m/s) | $v_L$ (m/s) | $k_e/k_D$ | $\lambda_e$ (Å) | $\sigma^2$ | $G_0$ (GPa) | $k_{BP}/k_e$ | $\lambda_{BP}$ (Å) | $k_{FSDP}/k_D$ | $k_D$ (Å$^{-1}$) |
|---|---|---|---|---|---|---|---|---|---|---|---|
| $SiO_2$ [4] | 2.21 [4] | 3746 [4] | 6058 [4] | 0.523 | 3.27 | 2.04 | 57.2 | 0.200 | 38.0 | 0.96 [4] | 1.58 |
| Densified $SiO_2$ (7.7 GPa, RT) | 2.24 [4] | 3744 [11] | 5902 [11] | 0.674 | 2.53 | 2.36 | 65.4 | 0.159 | 37.1 | 0.97 [4] | 1.59 |
| Densified $SiO_2$ (7.7 GPa, 600 °C) | 2.66 [4] | 4101 [11] | 6694 [11] | 0.751 | 2.14 | 1.69 | 73.0 | 0.193 | 25.8 | 1.07 [4] | 1.68 |
| Densified $SiO_2$ (7.7 GPa, 800 °C) | 2.68 [4] | 4180 [11] | 6921 [11] | 0.841 | 1.91 | 1.83 | 78.9 | 0.186 | 23.9 | 1.08 [4] | 1.68 |
| NS2 | 2.49 [12] | 3070 [12] | 5360 [12] | 0.745 | 2.22 | 1.63 | 37.7 | 0.206 | 25.0 | 1.34 [13] | 1.64 |
| NS3 | 2.43 [12] | 3159 [12] | 5347 [12] | 0.737 | 2.26 | 1.75 | 39.5 | 0.195 | 26.9 | 1.32 [13] | 1.63 |
| NS4 | 2.38 [12] | 3251 [12] | 5382 [12] | 0.703 | 2.38 | 1.87 | 43.0 | 0.187 | 29.6 | 1.17 [13] | 1.61 |
| $As_2S_3$ [14] | 3.14 [15] | 1418 [15] | 2625 [15] | 0.471 | 4.36 | 0.52 | 7.28 | 0.436 | 23.3 | 0.93 [16] | 1.32 |
| LiB8 | 1.95 [17] | 2489 | 4491 | 0.449 | 3.50 | 1.19 | 16.5 | 0.252 | 32.3 | 0.93 [18] | 1.72 |
| LiB14 | 2.05 [17] | 2915 | 5310 | 0.538 | 2.87 | 1.30 | 24.4 | 0.241 | 27.7 | 0.91 [18] | 1.75 |
| LiB22 | 2.16 [17] | 3388 | 6098 | 0.568 | 2.65 | 1.28 | 34.7 | 0.240 | 25.7 | 0.88 [18] | 1.79 |
| Glycerol [2] | 1.26 [19] | 1871 [19] | 3614 [19] | 0.451 | 3.16 | 0.82 | 5.43 | 0.375 | 19.6 | 0.79 [7] | 1.90 |
| Sorbitol [20] | 1.47 [21] | 2107 [20] | 4081 [20] | 0.345 | 4.01 | 0.42 | 7.29 | 0.490 | 19.0 | 0.74 [20] | 1.96 |
| PMMA [21] | 1.19 | 1420 [23] | 2780 [23] | 0.133 | 10.8 | 0.11 | 2.47 | 0.732 | 34.3 | 0.48 [23] | 1.88 |
| LJ glass [9] | 1.0 [9] | 3683 [9] | 9113 [9] | 0.238 | 2.91 | 0.80 | 16.1 | 0.364 | 16.0 | 1.81 [9] | 3.90 |
| Physical gel ($\rho$=0.7) [9] | 0.7 [9] | 2524 [9] | 4424 [9] | 0.079 | 9.89 | 0.04 | 4.52 | 0.858 | 26.8 | 0.19 [9] | 3.46 |
| Physical gel ($\rho$=0.5) [9] | 0.5 [9] | 2034 [9] | 3516 [9] | 0.059 | 14.8 | 0.01 | 2.08 | 0.937 | 36.7 | 0.16 [9] | 3.09 |

## Supplementary References